\documentclass[a4paper,usenatbib]{mnras}

\usepackage{newtxtext,newtxmath}

\usepackage[T1]{fontenc}
\usepackage{ae,aecompl}

\usepackage{graphicx}	
\usepackage{amsmath}	
\usepackage{amssymb}	

\usepackage{caption}

\usepackage{rotating}
\usepackage{lscape}
\usepackage{longtable}

\usepackage{subfigure}
\usepackage{graphicx}

\usepackage{dcolumn}
\newcolumntype{d}[1]{D{.}{.}{#1}}

\usepackage{multirow}
\usepackage[normalem]{ulem}

\usepackage{color}
\definecolor{red}{rgb}{1,0,0}
\definecolor{green}{rgb}{0,1,0}
\definecolor{blue}{rgb}{0,0,1}
\definecolor{sgm}{rgb}{1,0,1}

\newcommand*\rfrac[2]{{}^{#1}\!/_{#2}}

\title[Ionospheric characterisation over the MRO using the MWA]{Characterisation of the ionosphere above the Murchison Radio Observatory using the Murchison Widefield Array}

\author[C. H. Jordan et al.]{%
C.~H.~Jordan$^{1,2}$\thanks{E-mail: \href{mailto:christopher.jordan@curtin.edu.au}{christopher.jordan@curtin.edu.au}},
S.~Murray$^{1,2}$,
C.~M.~Trott$^{1,2}$,
R.~B.~Wayth$^{1,2}$,
D.~A.~Mitchell$^{2,3}$,
\newauthor
M.~Rahimi$^{2,4}$,
B.~Pindor$^{2,4}$,
P.~Procopio$^{2,4}$
and J.~Morgan$^{1,2}$
\\
$^{1}${International Centre for Radio Astronomy Research, Curtin University, Bentley, WA 6102, Australia}\\
$^{2}${ARC Centre of Excellence for All-sky Astrophysics (CAASTRO)}\\
$^{3}${CSIRO Astronomy and Space Science, Cnr Vimiera \& Pembroke Roads, Marsfield NSW, 2122, Australia}\\
$^{4}${School of Physics, The University of Melbourne, Parkville, VIC 3010, Australia}\\
}

\date{Last updated yyyy mmm dd; in original form yyyy mmmmmmmmm d}

\pubyear{2017}

\begin{document}
\label{firstpage}
\pagerange{\pageref{firstpage}--\pageref{lastpage}}
\maketitle

\newcolumntype{d}[1]{D{.}{.}{#1} }

\begin{abstract}
We detail new techniques for analysing ionospheric activity, using Epoch of Reionisation (EoR) datasets obtained with the Murchison Widefield Array (MWA), calibrated by the `Real-Time System' (RTS). Using the high spatial- and temporal-resolution information of the ionosphere provided by the RTS calibration solutions over 19 nights of observing, we find four distinct types of ionospheric activity, and have developed a metric to provide an `at a glance' value for data quality under differing ionospheric conditions. For each ionospheric type, we analyse variations of this metric as we reduce the number of pierce points, revealing that a modest number of pierce points is required to identify the intensity of ionospheric activity; it is possible to calibrate in real-time, providing continuous information of the phase screen. We also analyse temporal correlations, determine diffractive scales, examine the relative fractions of time occupied by various types of ionospheric activity, and detail a method to reconstruct the total electron content responsible for the ionospheric data we observe. These techniques have been developed to be instrument agnostic, useful for application on LOFAR and SKA-Low.
\end{abstract}

\begin{keywords}
atmospheric effects -- plasmas -- instrumentation: interferometers -- site testing
\end{keywords}


\section{Introduction}
In recent years, low-frequency astronomy research has taken significant leaps forward. Thanks to the steady increase in computing power over time, new low-frequency observatories have been created all around the world, many of which are motivated by uncharted territory in astronomical research, including the detection of a signal originating from the Epoch of Reionisation (EoR). These observatories include the LOw Frequency ARray (LOFAR; \citealt{vanhaarlem13}), the Donald C. Backer Precision Array for Probing the Epoch of Reionization (PAPER; \citealt{parsons14}), the Hydrogen Epoch of Reionization Array (HERA; \citealt{pober14}), and the Murchison Widefield Array (MWA; \citealt{tingay13}).

The Earth's ionosphere is known to affect low-frequency radio waves (typically less than 300\,MHz), but was largely ignored as an astronomical problem until these new low-frequency observatories came online. One large reason was that available instruments lacked the resolution to perceive these effects, which are now becoming apparent to new observatories. Now that the ionosphere acts as a significant obstacle to obtaining high-quality low-frequency observations, we must learn more about it.

Research into ionospheric calibration has already been conducted. For example, \citet{cotton04} detail a scheme for calibrating the 74\,MHz observations in the 10\,km `B' configuration of the Very Large Array. However, this scheme is limited to relatively compact interferometric arrays; the need for a general ionospheric calibration motivated the work by \citet{intema09}. Despite the availability of ionospheric calibration models, successful removal of ionospheric effects from low-frequency radio observations is difficult and computationally expensive. Is it instead possible to avoid data that features ionospheric activity? For an extremely sensitive project such as the EoR, quality assurance on data affected by the ionosphere may provide a more efficient means of detection. In addition, it is important to understand the variety of ionospheric effects that affect radio data, as well as the evolution of ionospheric activity over time, particularly for individual radio observatories.

In this paper, we focus our attention on ionospheric characterisation, analysing several nights worth of data and developing techniques for identifying adverse conditions above the Murchison Radio Observatory (MRO). In a companion paper (Trott et al. 2017, {\it in preparation}) we explore the impact of ionospheric activity on MWA EoR, and future Square Kilometre Array (SKA) EoR experiments. This complements previous studies of scintillation noise due to the ionosphere \citep{vedantham15,vedantham16}.

\subsection{The ionosphere}
\label{sec:ionosphere}
The ionosphere has been studied for many decades, typically with ionosondes, but also using global position system (GPS) satellites at coarse resolution (temporally 2 hours, spatially 5$^\circ$ and 2.5$^\circ$ in longitude and latitude, respectively; \citealt{arora15}). Such study has determined that the electron density of the ionosphere is most significant at altitudes between 150 and several thousand kilometres, although it peaks between 250 and 600\,km \citep{mannucci98}. \citet{erickson01} find that using only 4 GPS receivers, large-scale structures ($>$1000\,km) can be identified, but small-scale structures ($<$100\,km) require a high density of GPS receivers to provide an adequate number of pierce points. These small-scale structures can severely impact the quality of radio data obtained, but before we can hope to calibrate them, we must first identify and characterise them. Thus, the ionosphere has received attention as a substantial obstacle to low-frequency observations (typically $<$300\,MHz). However, it can also severely impact high-precision experiments at frequencies as high as 10\,GHz \citep{asaki07}.

The ionosphere is often described by its total electron content (TEC), which is a column-density of electrons integrated along a line through the plasma. These electrons act to refract the incoming, planar wavefronts of far-field sources, inducing a delay that is observed as a change in phase. Because the MWA is an interferometer, measuring the phase differences across baselines, it is unable to identify the total amount of phase contributed by the TEC. Instead, the MWA measures the differences in these phases, caused by variations of the TEC.

Variations of the TEC can be categorised into groups; for example, an interferometer will detect a first-order variation of the TEC as a slope across the whole field of view. This will manifest a bulk offset apparent to all observed radio sources, and is simple to calibrate. Higher-order variations in the TEC are much more difficult to calibrate; not only do radio sources have the potential to scintillate (focus/de-focus, change apparent flux density, visibly distort etc.), but the responsible anisotropies form a small fraction of the total TEC. Moving into the SKA era, it is pertinent to understand these higher-order TEC variations; how many radio sources are required to identify and calibrate small-scale ionospheric structures? How quickly does the ionosphere change? These questions comprise a large motivation for this work.

When observing through the ionosphere at an arbitrary elevation and azimuth, the TEC modulating incoming radio waves is referred to as the slant TEC (STEC). The apparent spatial offsets of sources due to ionospheric refraction is given by:

\begin{equation}
  \label{eq:stec}
  \Delta\theta \simeq - \frac{1}{8\pi^2} \frac{e^2}{\epsilon_0 m_e} \frac{1}{\nu^2} \nabla\text{STEC [rad]}
\end{equation}\\
where $e$ is the electron charge and $m_e$ is the electron mass. For convenience, in this work we use the terms `ionosphere' and `STEC' interchangeably. As seen in Equation~\ref{eq:stec}, the refractive shift due to the ionosphere is proportional to $\lambda^2$; this implies that low-frequency radio observations can be severely impacted by adverse ionospheric activity. If high-sensitivity and high dynamic range experiments are ever going to be realised, such as detection of the EoR signature, diagnosis and assessment of our ionosphere are required.

The MWA is an SKA-Low precursor, comprised of a 128-element dipole array capable of observing frequencies between 80 and 300\,MHz. The MWA has demonstrated the ability to characterise ionospheric activity by revealing prominent tubular structures aligned with the Earth's magnetic field \citep{loi15a}. In a similar way, this work aims to use the MWA to characterise ionospheric activity in the context of EoR datasets.

\section{Observations and calibration}
MWA EoR observations are conducted in a number of different fields, selected primarily to be in regions of low-sky temperature at high Galactic latitude. In this work, all observations were conducted on the `EoR-0' field ($\alpha = 0^\text{h}$, $\delta = -27^\circ$) \citep{jacobs16}. A total of 927~observations were used from September to December 2015, each comprising 112~seconds of data, culminating in just under 29\,hours of data. Observations are referenced by their observational number (`obsid'), which is a timestamp using the global positioning system format in seconds. Table~\ref{tab:obsids} shows all observations used in this work.

\begin{table}
  \begin{center}
    \caption{Dates of observations used in this work. Each observation (`obsid') represents a timestamp in the global positioning system format, and was calibrated by the RTS using 1000 sources in the EoR-0 field.}
    \label{tab:obsids}
    \begin{tabular}{ ccc }
      \hline
      Date & First obsid & Final obsid \\
      \hline
      2015-09-08 & 1125762376 & 1125770552 \\
      2015-09-10 & 1125934704 & 1125942880 \\
      2015-09-12 & 1126107032 & 1126115208 \\
      2015-09-14 & 1126279360 & 1126287536 \\
      2015-09-22 & 1126971240 & 1126976848 \\
      2015-09-24 & 1127141000 & 1127149176 \\
      2015-10-06 & 1128174968 & 1128183144 \\
      2015-10-08 & 1128347304 & 1128355472 \\
      2015-10-10 & 1128519632 & 1128527800 \\
      2015-10-12 & 1128698424 & 1128700128 \\
      2015-10-14 & 1128864288 & 1128872456 \\
      2015-10-20 & 1129381272 & 1129389440 \\
      2015-10-22 & 1129553600 & 1129561160 \\
      2015-11-23 & 1132312496 & 1132318960 \\
      2015-11-29 & 1132831192 & 1132836064 \\
      2015-12-03 & 1133176976 & 1133180752 \\
      2015-12-05 & 1133349792 & 1133353080 \\
      2015-12-09 & 1133695544 & 1133697736 \\
      2015-12-11 & 1133868480 & 1133868728 \\
      \hline
    \end{tabular}
  \end{center}
\end{table}

MWA observations provide 30.72\,MHz of total bandwidth over 24 coarse channels. The correlator output is provided every 2\,s with a frequency resolution of 40\,kHz, and is written as a discrete file on a 112\,s cadence. For these observations, the centre frequency was 182\,MHz, and calibration was performed every 8\,s to provide 14 `snapshot' intervals per observation \citep{bowman13}.

\subsection{Measuring ionospheric effects with the RTS}
\label{sec:rts}
As its name suggests, the RTS is designed for real-time calibration via a calibrator measurement loop \citep{mitchell08}. Note that in this context, a calibrator refers to a radio source, typically an unresolved radio galaxy. Ionospheric diagnosis also forms a key part of this calibration loop. For our purposes, a large number of pierce points (order 1000) is useful for ionospheric characterisation; however, running the RTS with this many calibrators is too computationally expensive for real-time analysis. Fortunately, for the purposes of EoR datasets, real-time analysis is not required, allowing thorough off-line calibration. A summary of this calibrator measurement loop is as follows \citep[for more detail see][]{mitchell08}. The loop runs after standard direction-independent gain and bandpass calibration.

(i) From a large catalogue of radio sources and their flux densities, visible sources are ranked according to their expected received power, attenuated by the antenna primary beam;

(ii) Calibrators have their contributions in visibility space subtracted (pre-peeled), to reduce their sidelobe contribution as much as possible;

(iii) Begin a loop over every calibrator, starting with the brightest. For this calibrator, restore its subtracted visibilities in step (ii), and rotate the visibility space such that the expected position of this calibrator is in the phase centre. Average the visibilities of the calibrator in time (typically 8\,s) and within each frequency channel (of width 1.28 MHz) to reduce thermal noise;

(iv) A single $\lambda^2$ dependence of ionospheric refraction is fitted to all coarse channels (over a total of 30.72 MHz). To first order the small refractive phases result in residual visibilities with non-zero imaginary parts that scale linearly with flux density, $u$, $v$ and $\lambda^2$, and this system is solved via linear least squares to model the refraction. Re-rotate the phase centre to the apparent position of the calibrator;

(v) For a small number of the brightest calibrators, Jones matrices can also be estimated, representing the polarised voltage gain of each antenna;

(vi) If the gain and ionospheric measurements pass certain goodness-of-fit tests, they are used to peel the visibilities of the calibrator from the measurement set. If the tests are not passed, the original subtraction in step (ii) is reapplied. For every other calibrator to be processed, repeat from step (iii).

It is this procedure that facilitates the work presented in this paper, providing high spatial- and temporal-resolution measurements of the ionosphere over a large field of view. As discussed in Section~\ref{sec:ionosphere}, the ionosphere is comprised of various order variations in the TEC. Most of the time, the TEC is approximately constant across large solid angles of the Earth. The higher-order variations of the TEC, which the RTS probes with exquisite resolution, form only a small fraction of the total TEC. As typical measurements of the TEC are restricted to GPS satellites, the RTS is well positioned to study these higher-order variations.

Throughout, we associate deviations from catalogue positions measured with the RTS with ionospheric refraction. The number of sources used for RTS calibration can be specified at runtime, although typically either 300 or 1000 sources are used. In this work, all observations have been calibrated with 1000 sources. These sources were derived from catalogues provided by the Positional Update and Matching Algorithm (PUMA; \citealt{line17}). Further details of RTS design and calibration can be found in \citet{mitchell08}. If the number of calibrator sources used is small enough (100), it is possible to run the RTS in real-time. Such an operation would provide a continuous, high-resolution characterisation of the ionosphere.

As the ionospheric refraction fits also provide amplitude estimates for every calibrator source, it is possible to investigate the scintillation effects of the ionosphere; however, this paper focuses only on the difference between apparent and expected positions of calibrator sources (ionospheric offsets). Additional higher-order effects of the ionosphere may be investigated in future publications.

\section{Ionospheric analysis with \textsc{cthulhu}}
Upon running the RTS, status information along with measured ionospheric corrections are dumped into log files. To extract and analyse the information pertaining to the ionosphere, we have designed a software suite in the \textsc{Python} programming language, titled \textsc{cthulhu}. A summary of the functions of \textsc{cthulhu} is as follows:

(i) Scraping of RTS calibration logs for observational metadata, such as the observational number, pointing centre and observing frequency, as well as ionospheric corrections for every calibrator source. These corrections are typically listed every 8\,s over 112\,s for EoR data;

(ii) Calibrator positions are listed in Alt.-Az. co-ordinates, while ionospheric corrections are listed as $l$ and $m$ direction cosines. These are converted to have both calibrator positions and ionospheric corrections in RA-Dec. co-ordinates;

(iii) Calculation of a metric describing ionospheric quality for observations (the metric is detailed in Section~\ref{sec:metric});

(iv) Optionally, the ionospheric corrections may be used to reconstruct a scalar field representing the TEC (detailed in Appendix~\ref{app:reconstruction}). Plotting functions are provided to generate diagnostic plots, including two-dimensional power spectra derived from the reconstructed TEC.

The source code of \textsc{cthulhu}, as well as examples and sample data, are provided on GitHub\footnote{\url{https://github.com/cjordan/cthulhu}}. \textsc{cthulhu} has been designed to be as modular as possible, such that future modification would be easily accomplished; for example, additional co-ordinate systems rather than RA-Dec. may be implemented.

\section{Results and discussion}
\subsection{Metric of ionospheric activity}
\label{sec:metric}
Using a range of statistics, we have determined a useful formula for quickly scoring a dataset for the presence and class of ionospheric activity. There are two properties of ionospheric activity that we wish to capture in these statistics; firstly, the overall strength of the activity, and secondly, the level of structure in the activity (e.g. whether it is highly anisotropic or turbulent on some typical scale). Focusing on the first property, the statistics investigated include:

\begin{enumerate}
\item The median magnitude of ionospheric offsets;
\item The standard deviation, skewness and kurtosis of the reconstructed scalar-field TEC;
\end{enumerate} 
while those focusing on the second property include:
\begin{enumerate}
\setcounter{enumi}{2}
\item The standard deviation of the Laplacian and Hessian over the reconstructed scalar-field TEC;
\item The dominant normalised eigenvalue determined by a principal component analysis (PCA). The PCA is performed directly on the source offsets, with the offset in both orthogonal directions forming the 2-dimensional space of the PCA. The fraction of the total variance explained by a given component is equivalent to its normalised eigenvalue, $\epsilon_i/\sum \epsilon_i$. Because the data are two-dimensional, the principal component explains between 50 and 100 per cent of the variance, which quantifies how well the data are reproduced with only a single dimension.
\end{enumerate}

As could be expected, each of these statistics are highly correlated, with those within each group being more highly correlated with each other. Thus for simplicity and efficiency, we chose a representative statistic from each group - (i) the median offset, and (ii) the dominant normalised PCA eigenvalue - as complementary descriptions of the data. The median offset is sensitive to the amplitude of the gradient of the STEC, but completely insensitive to directionality, including anisotropy. Conversely, the dominant normalised PCA eigenvalue is sensitive to the level of anisotropy, but completely insensitive to the amplitude of the offsets.

\begin{figure*}
  \includegraphics[width=\textwidth]{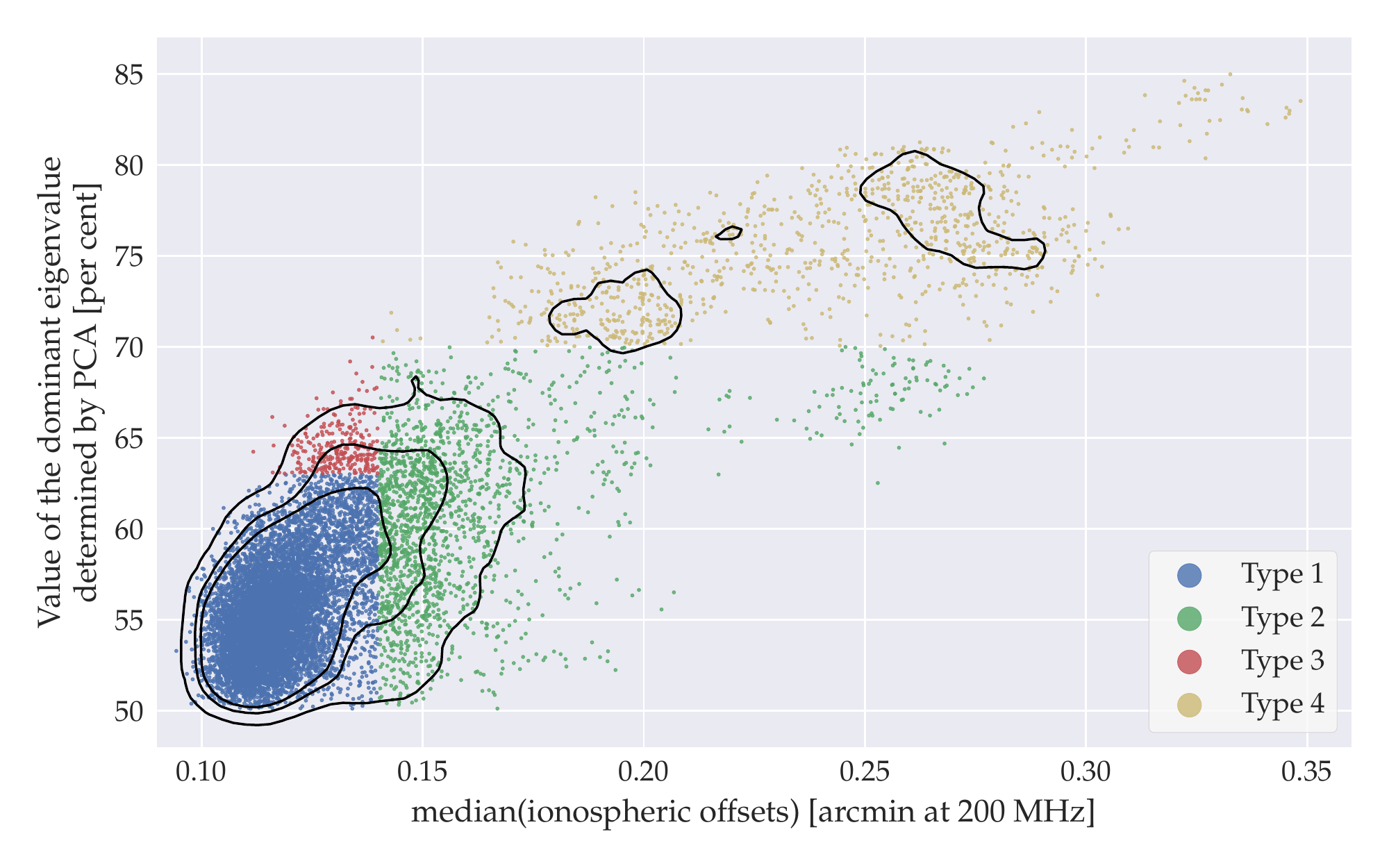}
  \caption{Scatter plot for the dominant eigenvalue determined by PCA vs. median ionospheric offset for each observation. Each of the types of ionospheric activity described in Section~\ref{sec:metric} are highlighted here; examples of each type can be seen in Fig.~\ref{fig:tecs}. The fractional size of each population is approximately 74, 15, 2.3 and 8.4 per cent for types 1 through 4, respectively. The contour levels indicate the density of observations, and are at 90, 60 and 30 per cent.}
  \label{fig:metric_scatter}
\end{figure*}

A scatter plot of the dominant eigenvalue determined by PCA vs. median ionospheric offsets can be seen in Fig.~\ref{fig:metric_scatter}. The ionospheric data from each observation were inspected, and in conjunction with this plot, four populations have been highlighted, qualitatively described as:

(i) weakly-correlated ionospheric offset directions, with small-magnitude offsets. These observations appear to have very little ionospheric activity;

(ii) weakly- or moderately-correlated ionospheric offset directions, with moderate- or large-magnitude offsets. It is possible that some instances of this kind of ionospheric activity show a strong preference in the direction of the ionospheric offsets, but our field of view is too restricted to see this. These may be good examples of TIDs and isotropic turbulence;

(iii) highly-correlated ionospheric offset directions, with weak-magnitude offsets. These types of ionospheres appear to be largely inactive, but contain weak coherent structure, possibly due to increased geomagnetic activity; and

(iv) highly-correlated ionospheric offset directions, with large-magnitude offsets. These observations are similar to those of \citet{loi15a}, and feature extreme ionospheric activity. This type of ionospheric activity was only witnessed twice in our nineteen nights of observing, suggesting that this behaviour is uncommon.

We have labelled each of these classifications as types 1 through 4. These groups have been specifically classified by the following criteria:

\begin{align*}
  m < 0.14 &\text{ and } p < 63 \Rightarrow \text{Type 1}\\
  m > 0.14 &\text{ and } p < 70 \Rightarrow \text{Type 2}\\
  m < 0.14 &\text{ and } p > 63 \Rightarrow \text{Type 3}\\
  m > 0.14 &\text{ and } p > 70 \Rightarrow \text{Type 4,}
\end{align*}
where $m$ is the median ionospheric offset across all sources in arcmin at 200\,MHz, and $p$ is the dominant eigenvalue determined by PCA in per cent. These criteria divide the total population into approximately 74, 15, 2.3 and 8.4 per cent for ionospheric types 1 through 4, respectively.

\begin{figure*}
  \includegraphics[width=\textwidth]{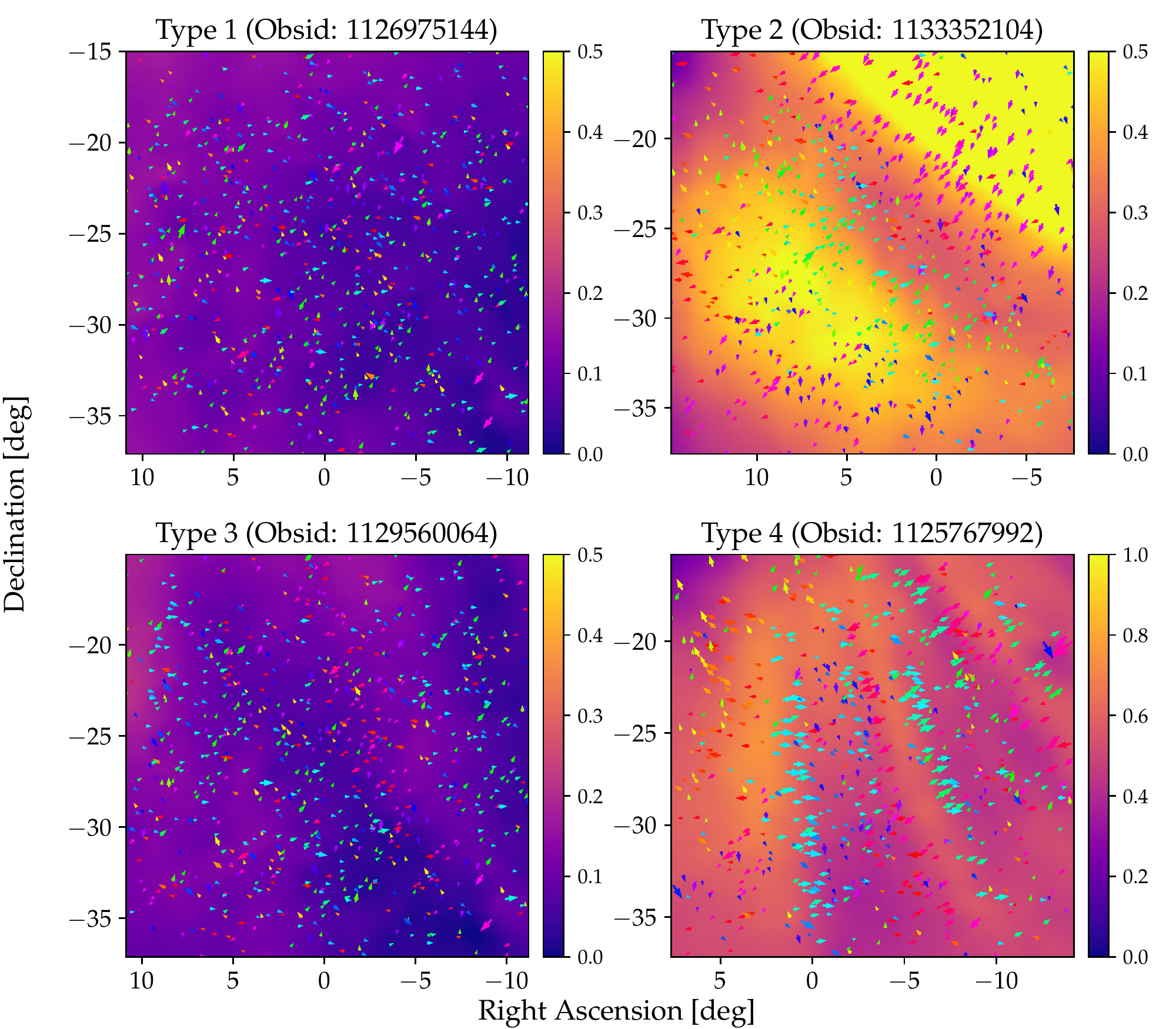}
  \caption{Four observations with their ionospheric offsets overlaid on their corresponding reconstructed TEC scalar fields. The units of the colour-scale for each plot are TECU (or 10$^{16}$\,m$^{-2}$). Each ionospheric offset is colour-coded according to its direction, and is scaled by a factor of 60. The reconstructed TECs have had their minimum values subtracted, to remove the arbitrary constant of integration. Note that ionospheric offsets are derived from an observing frequency of 200\,MHz, and we assume a height of 400\,km in order to calculate the TEC units.}
  \label{fig:tecs}
\end{figure*}

For the remainder of this paper, we refer back to these classifications as designated types of ionospheric activity. Fig.~\ref{fig:tecs} presents an example of each of these four types. While type 1 ionospheric conditions can be assumed to be best for radio observing (and similarly, type 4 the poorest), at present, it is uncertain which of type 2 or 3 conditions are poorer. It is beyond the scope of this paper to determine which is worse, and follow-on work from this paper which analyses the effect of the ionosphere on EoR data will resolve the difference. However, we have proceeded under the assumption that the quality of radio data degrades as the ionospheric type increases, as anisotropic distributions seen in type 3 data (such as what is shown in Fig.~\ref{fig:tecs}) may affect only some \emph{uv} frequencies, while type 2 data is more likely to affect all frequencies, and may be easier to calibrate given their large-scale structures. The possible \emph{uv}-plane bias introduced by type 3 ionospheric conditions would be particularly poor for EoR power spectra, with additional power introduced into some modes but not others.

\begin{figure}
  \includegraphics[width=\columnwidth]{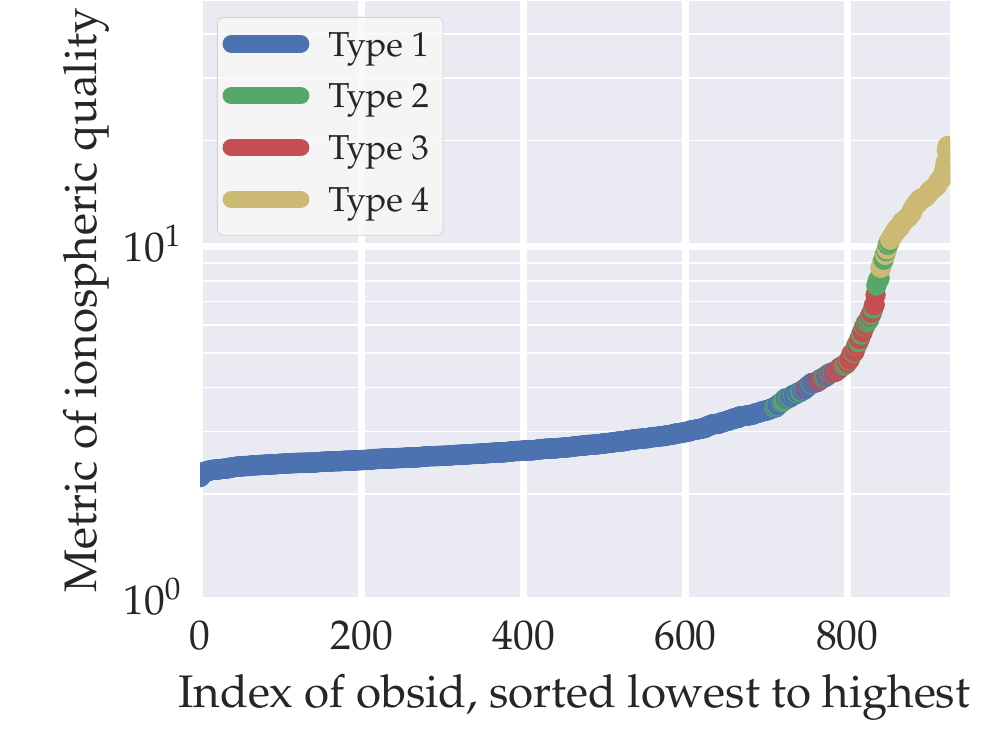}
  \caption{Plot of the ionospheric quality metric for every observation used in this work, sorted from lowest to highest. Sections of the line highlight the type of ionosphere determined for that observation.}
  \label{fig:metric_sorted}
\end{figure}

We have constructed a metric using the statistics described above. The metric is designed such that larger values designate poorer observing conditions, and is defined as:

\begin{equation*}
  \label{eq:metric}
  M =
  \begin{cases}
    25m + 64p\left( p-0.6 \right),& \text{if } p > 0.6\\
    25m,                          & \text{otherwise}\\
  \end{cases}
\end{equation*}\\
where $M$ is the metric of ionospheric quality. This places a quadratic dependence on eigenvalue determined by PCA, biasing this statistic for strongly-aligned ionospheric offsets. Fig.~\ref{fig:metric_sorted} shows the metric value of each observation, sorted from lowest to highest, to indicate the fractions of active and inactive ionospheres. Type 2 ionospheres start to blend with Type 1 ionospheres around index 650; if we consider this as point discriminating active and inactive ionospheres, then 69 per cent of observations lack ionospheric activity.

\subsection{Spatial structure}
\citet{mevius16} used the LOFAR radio telescope to characterise ionospheric structure by tracking a single source over a large amount of time. This statistic allows an understanding of the spatial correlation lengths by tracing the celestial-frame source as it pierces a track through the ionosphere. They diagnose spatial correlation in terms of the phase structure function:

\begin{equation*}
  D(r) = \left< \left( \phi(r') - \phi(r'+r) \right)^2 \right>
\end{equation*}\\
where $r$ is the baseline length, $\phi(r)$ is the phase at baseline length $r$, and $D(r)$ is the variance. In the work described, the ensemble average is conducted over the temporal phase of the source of each LOFAR baseline. It is possible to conduct this type of analysis with our MWA observations, with the addition of using many sources rather than a single one. With the high spatial and temporal resolution of the MWA data, and the co-temporal phase estimation of many sources, the ensemble average may be conducted over either dimension. In the limit of the frozen flow model \citep{vandertol07}, both approaches should yield the same structure function. Deviations from this outcome may indicate the non-stationary nature of ionospheric structure over the wide MWA field-of-view. In the following subsections, we discuss both approaches, but firstly define the reconstructed phase, $\phi$, on a baseline $(u,v)$, as:

\begin{equation*}
  \phi = 2\pi(u\Delta{l} + v\Delta{m}),
\end{equation*}\\
where $\Delta{l}$, $\Delta{m}$ denote the measured source offset vector components.

\subsubsection{Phase variance of a spatial ensemble}
\begin{figure}
  \includegraphics[width=0.45\textwidth]{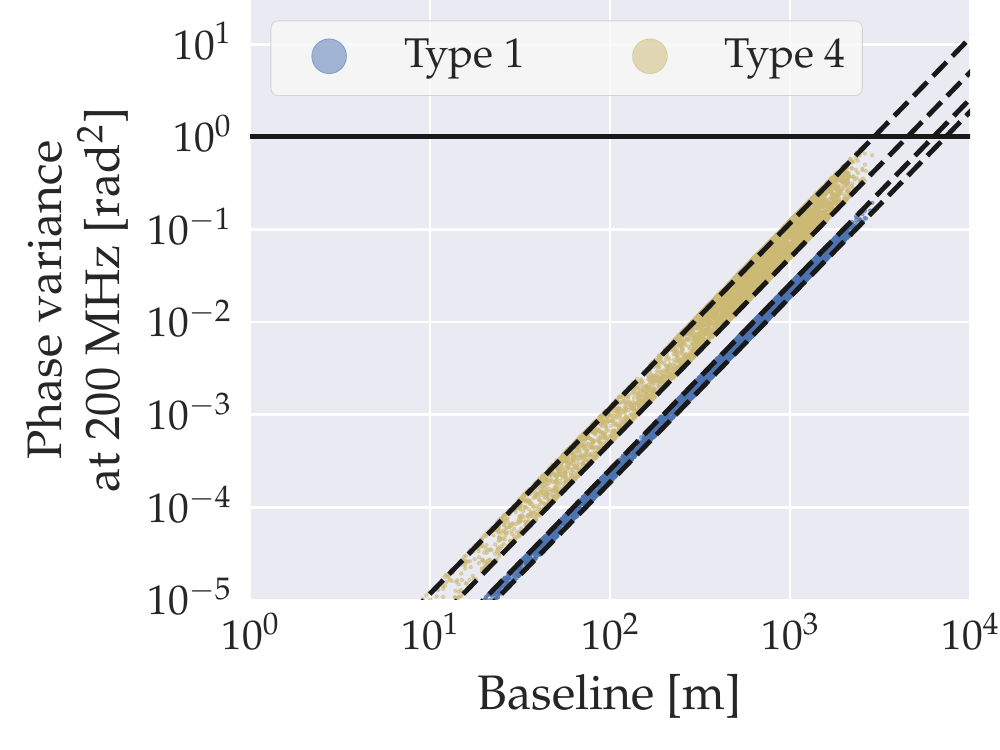}
  \caption{Diffractive scale estimation of Type 1 and 4 observations, using a spatial ensemble average to determine phase variance for a given baseline. Extrapolated values estimate diffractive scales of $r_{\rm diff} = 6.3-7.3$\,km and $r_{\rm diff} = 2.9-4.5$\,km for types 1 and 4, respectively.}
  \label{fig:rdiff_ensemble}
\end{figure}

The phase variance is computed across all sources used for ionospheric pierce points, representing the structure of the approximate 25-by-25~degree field-of-view of the MWA. This mode takes a snapshot view of the ionosphere above the array. Fig.~\ref{fig:rdiff_ensemble} displays reconstructed phase variance estimates across all 8128 baselines of the MWA for Type 1 and 4 ionospheres, while their estimated scales are $r_{\rm diff} = 6.3-7.3$\,km and $r_{\rm diff} = 2.9-4.5$\,km, respectively.

Due to the large field of view of the MWA, it is possible that this analysis exacerbates differences between STEC (which we measure) and VTEC (which is elevation dependent). When performing our spatial ensemble analysis, we considered this point, but found no significant effect that would cause a bias in the phase variance.

\subsubsection{Phase variance of a temporal ensemble}
\begin{figure}
  \includegraphics[width=0.45\textwidth]{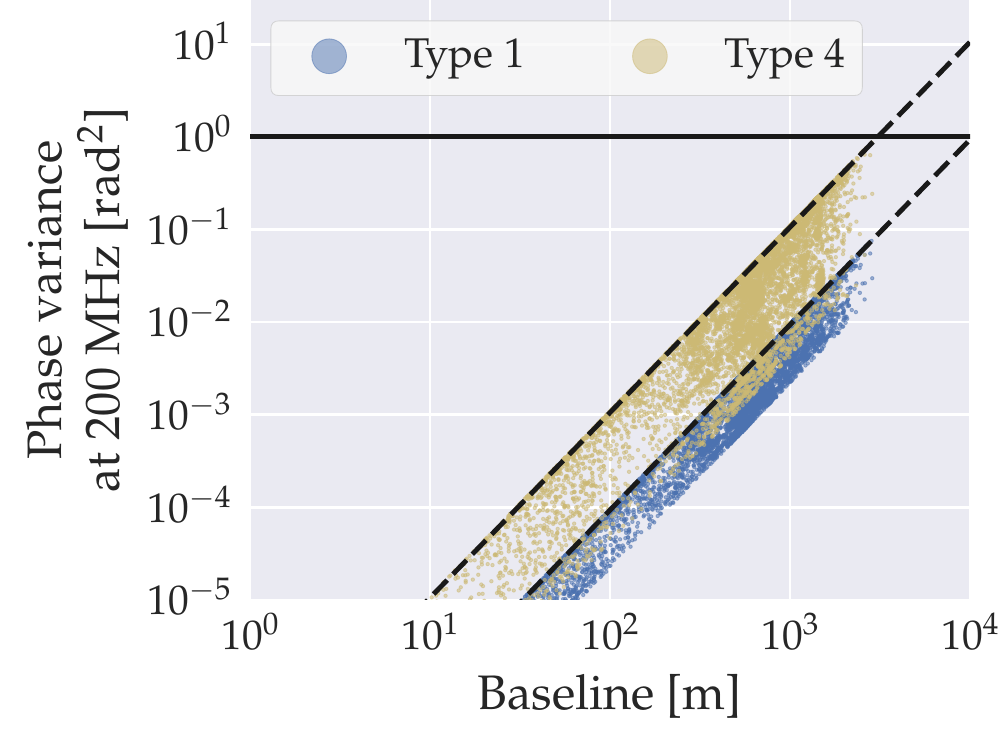}
  \caption{Diffractive scale estimation of Type 1 and 4 observations, using a temporal ensemble average of an indicative source to determine phase variance for a given baseline. Extrapolated values estimate diffractive scales of $r_{\rm diff} > 10$\,km and $r_{\rm diff} > 3.1$\,km for types 1 and 4, respectively.}
  \label{fig:rdiff_source}
\end{figure}

If the conditions of the ionosphere vary spatially, particularly over the large field-of-view probed instantaneously by the MWA, then a spatial ensemble average may over-estimate the phase variance, thereby under-estimating the diffractive scale. If we follow the methods of \citet{mevius16} to track a single, bright calibrator temporally, then we probe a more contained region of the ionosphere. Fig.~\ref{fig:rdiff_source} displays the reconstructed phase variance estimates for Type 1 and 4 ionospheres. The corresponding estimated scales are $r_{\rm diff} > 10$\,km and $r_{\rm diff} > 3.1$\,km for types 1 and 4, respectively. The anisotropy is evident in the Type 4 data, where baseline vectors perpendicular to the principal component yield very small variance estimates.

By construction, this analysis with MWA baselines utilises a power-law index of 2. However, pure Kolmogorov turbulence has an index of $\rfrac{5}{3}$, and \citet{mevius16} measure an average index of 1.89. To determine the diffractive scales with a slope that is purely turbulent for these data, we can use the phase variance at the average baseline length of the MWA (2.2\,km) and extrapolate using a slope of $\rfrac{5}{3}$. This places the upper limits of $r_{\rm diff}$ to be $32$ and $27$\,km for Type 1 and 4 data, respectively.

When the source offsets are highly anisotropic, the diffractive scale does not have a physical definition that can be associated with a turbulent scale size, and the two methods yield comparable results. When the data show turbulent-like structure, the temporal average yields larger diffractive scales, due to an irregular structure in the ionosphere. From these results, we suggest that the typical diffractive scales of extremely active ionospheric data is approximately a few kilometers, while inactive ionospheric data has a diffractive scale on the order of ten kilometers. The temporal-average scales reported by \citet{mevius16} using LOFAR are broadly consistent with ours, after converting their results from a 150\,MHz basis to our 200\,MHz.

\subsection{Temporal properties}
\label{sec:temporal}
At present, the temporal behaviour of the ionosphere is not well characterised. In order to calibrate their instruments, low-frequency radio astronomers are interested in coherence-time lengths of ionospheric activity, as well as some forecasting ability for the remaining night-time hours. Fortunately, using EoR datasets, we are able to analyse temporal ionospheric variations over many hours.

\subsubsection{Celestial frame variation}
\label{sec:celestial}
Temporal ionospheric coherence may be examined by application of the two-point correlation function:

\begin{equation*}
  \label{eq:two_point}
  \rho(t) = \frac{\Big\langle \big( O(t) - \big\langle O(t) \big\rangle \big) \big( O(t + \Delta t) - \big\langle O(t + \Delta t) \big\rangle \big) \Big\rangle}{\sigma(t)\sigma(t + \Delta t)}
\end{equation*}\\
where $O(t)$ represents the ionospheric offsets at the observation time $t$, and the ensemble average is performed spatially across pixels in the reconstructed STEC. For this analysis, we use only the ionospheric offsets in the $l$-direction; note that the $l$-direction is typically orthogonal to ionospheric structures seen in Type 3 and 4 data.

It should be noted that the EoR datasets used in this work have used a `drift-and-shift' observing strategy: over the course of the observations, the pointing centre is periodically changed to keep the EoR-0 field close to the centre of the primary beam. While it is possible to analyse temporal correlations of the ionosphere across an entire evening, they will be affected by systematic effects, such as a direction-dependent primary beam, as well as introducing different parts of the ionosphere for every `shift'. Thus, because of the drift-and-shift observing mode, we must analyse observations available between adjusted pointing centres if we wish to avoid bias.

\begin{figure*}
  \includegraphics[width=\textwidth]{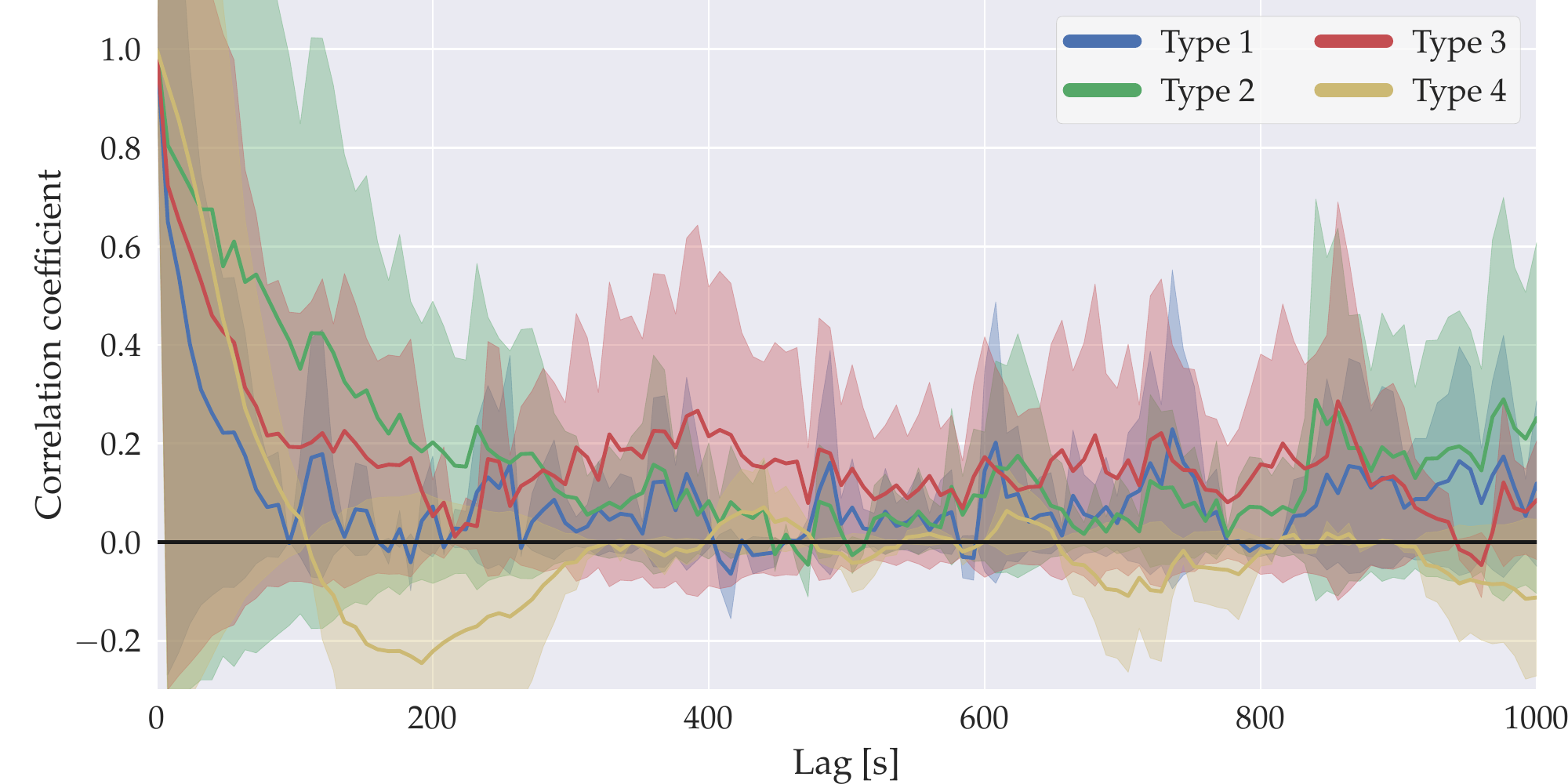}
  \caption{Temporal correlations for each of the four ionospheric types discussed in this work. The solid lines represent the ensemble averaged values, while the shaded region represents the error. A solid black line has been added at $\rho = 0$ for clarity. Type 1 decorrelates rapidly due to noise-like data, which in turn indicates a lack of ionospheric presence. Type 4 data smoothly decorrelates into negative correlation coefficient values, which is indicative of regular structures in the ionosphere.}
  \label{fig:temporal}
\end{figure*}

Fig.~\ref{fig:temporal} shows this celestial-frame analysis applied to each of the four designated types of ionospheric activity. The observation ranges used within this plot are 1126972456 to 1126974168 for Type 1, 1133350400 to 1133352104 for Type 2, 1129559208 to 1129560920 for Type 3 and 1125764328 to 1125766040 for Type 4. Each of these observation ranges are within a `shift' of the observing mode, to mitigate systematic biases. The errors in Fig.~\ref{fig:temporal} are determined with:

\begin{equation*}
  E(t) = \sqrt{2} \frac{ \left< \rho(t) \right> }{\sqrt{N(t)}} 
\end{equation*}\\
where $N(t)$ is the number of data points available at a lag $t$.

Type 1 ionospheric data decorrelates most rapidly of all types, and has correlation coefficients consistent with noise after approximately 100\,s. This indicates that Type 1 data are noise-like, most likely because any ionospheric refractions are small and incoherent. \citet{vedantham15} find that pure Kolmogorov-type turbulence has temporal coherence only on small timescales for short baselines, which may be responsible for the results we describe here.

Type 2 ionospheric data decorrelates least rapidly of all types, likely due to the large-scale structures present in the observations; see Fig.~\ref{fig:tecs}. Fig.~\ref{fig:temporal} is thus providing an indication of the scale of the ionospheric structures. After approximately 400\,s, the Type 2 correlation coefficients are noise-like, which also indicates that the ionospheric structures are not regular, or are not orthogonal to the sidereal rate of rotation of the celestial sources used.

Type 3 ionospheric data has perhaps the least noise-like correlation coefficients across the 1000\,s of observations. Like the Type 2 data used, compared to the field of view, the ionospheric structures appear to be relatively large and regular, but little if any ionospheric structure is present between the `valleys'.

Finally, Type 4 ionospheric data smoothly varies from lag values between 0 and 300\,s. As the Type 4 data is regular, one might expect celestial sources to refract in opposite directions as they `travel across' the ionospheric structures; this seems to manifest visibly in our temporal correlation plot. As with the Type 2 data, the rate of decrease in the correlation coefficients hint at the size of the ionospheric structures present.

\citet{intema09} discuss the temporal resolution required to calibrate ionospheric structures; while a high-temporal-resolution view of the ionosphere may be required to precisely calibrate the induced phase changes in observed radio data, the results presented in Fig.~\ref{fig:temporal} indicate that the majority of ionospheric structure changes slowly with time. Thus, because the ionosphere generally does not change dramatically on small ($<$50s) timescales, it is possible that the majority of ionospheric structure could be calibrated at the temporal resolution used by these data (8s).

\subsubsection{Geocentric frame variation}
\label{sec:geocentric}
By analysing ionospheric activity in a geocentric frame, we can eliminate effects of the Earth's sidereal rate of rotation. Geocentric reference frames can be obtained by converting celestial source positions and their ionospheric offsets into an $(l,m)$ co-ordinate system, using the local zenith as the origin. This method has the advantage of avoiding complications in spherical co-ordinate systems, such as Alt.-Az., where sources are spread across a wide range of latitudes. Here, to analyse the temporal properties of the ionosphere, we use the reconstructed TEC scalar fields projected in a single geocentric frame.

\begin{figure}
  \includegraphics[width=\columnwidth]{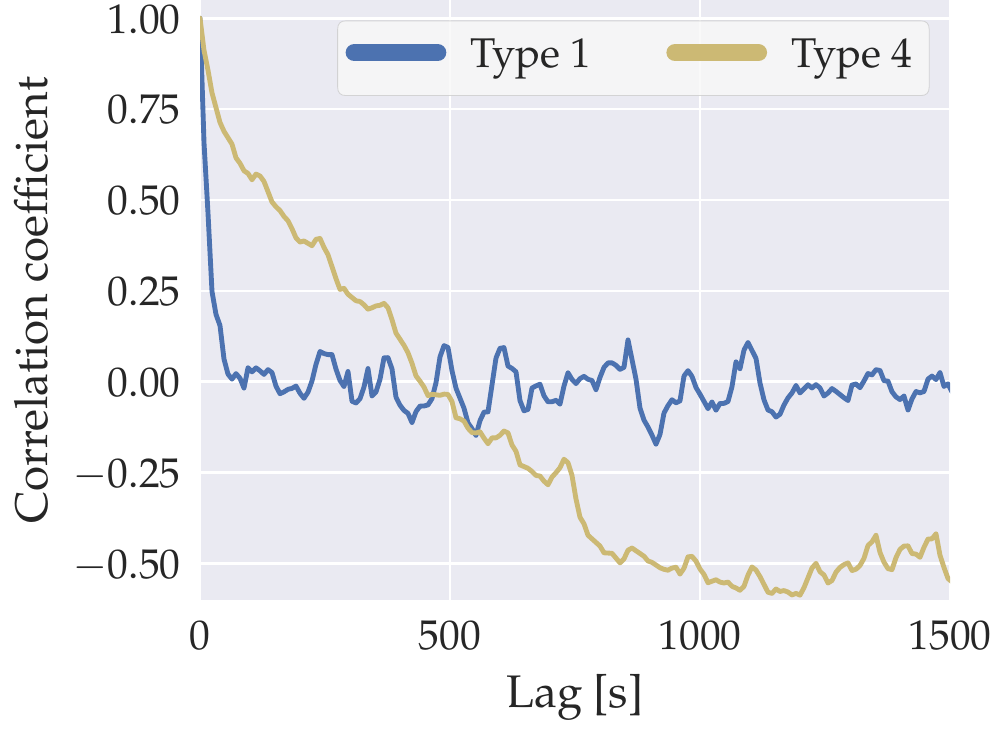}
  \caption{Temporal two-point correlation using 1500 seconds of Type 1 and 4 observations. These data make use of reconstructed TECs in a geocentric frame. Type 1 data decorrelates rapidly, due to its lack of significant ionospheric structure. Type 4 data decorrelates slowly and smoothly toward $\rho \approx -$~0.5; this indicates that the ionospheric structures seen in Type 4 data are slowly moving in a geocentric frame, and as they appear regular (similar to a two-dimensional sinusoid), negatively correlate given enough time.}
  \label{fig:temporal_geocentric}
\end{figure}

Fig.~\ref{fig:temporal_geocentric} displays the two-point correlation function computed over these re-projected TEC scalar fields. Type 1 data completely decorrelates within approximately 50\,s; similar to Section~\ref{sec:celestial}, we attribute this result to noise-like data, which indicates a lack of ionospheric activity. Also similar to Section~\ref{sec:celestial}, Type 4 data decorrelates smoothly down to $\rho \approx -$~0.5 at a lag of approximately 1000\,s. This slow decorrelation manifests as a result of the ionospheric structures slowly moving across the field of view, and negatively correlating due to their regular, sinusoid-like appearance.

Future work using this analysis will be useful to better understand subtle ionospheric effects; for example, by using many more pierce points, we may reconstruct TECs with higher precision, and better determine the variability of the ionospheric structures seen in Type 4 data, or identify weak structures present in data with seemingly inactive ionospheric activity.

\subsubsection{Metric variation}
In the other temporal-focused sections of this paper, we have determined that ionospheric activity tends to persist over long spans of time. It is prudent to analyse the variation of our metric of ionospheric quality over an observation session; if the variance of the metric is significant within a short span of time, then it may be misleading to an observer that would otherwise assume the ionosphere is behaving favourably for their low-frequency instrument.
Many low-frequency projects, including EoR observations with the MWA, only observe in the evening; it is useful to measure the activity of the ionosphere over the course of an evening to understand typical variations.

\begin{figure}
  \includegraphics[width=\columnwidth]{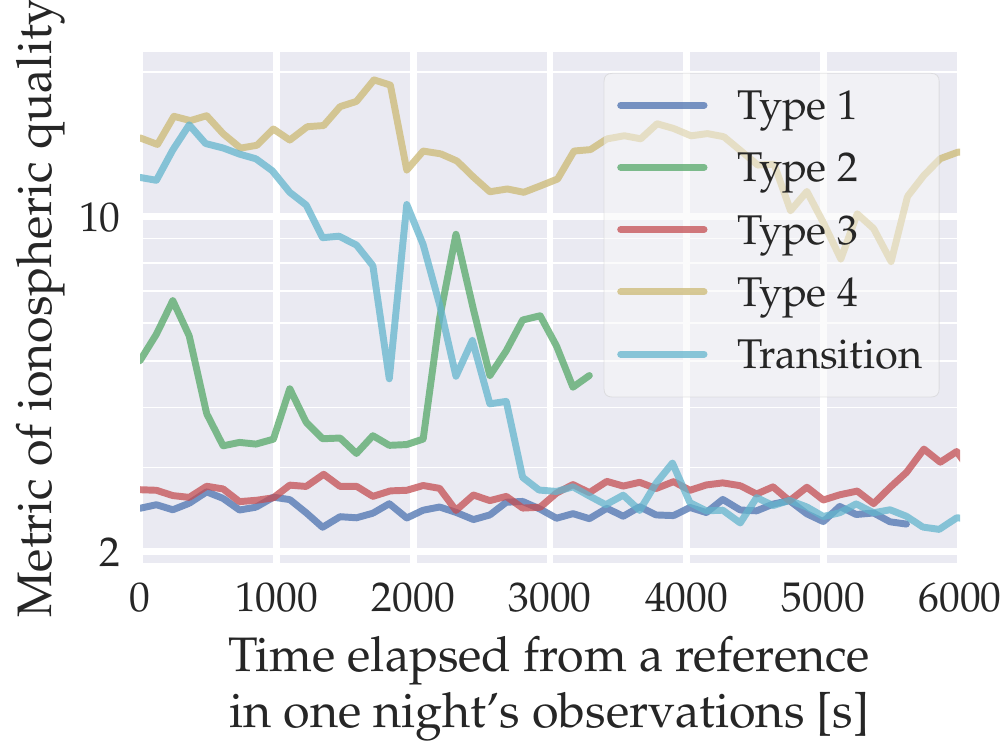}
  \caption{Line plot of the ionospheric quality metric for each of the four ionospheric types discussed in this work, as well as another night showing a transition. Types 1, 3 and 4 appear to have little variation in their metric values over the time series, whereas Type 2 is less stable. The transitional observations start active and become inactive after almost an hour; this indicates that ionospheric activity should be monitored at least hourly for observational quality assurance.}
  \label{fig:metric_variation_time}
\end{figure}

Fig.~\ref{fig:metric_variation_time} shows the metric of ionospheric quality over five tracks of time, one for each ionospheric type and another night which transitions from Type 4 to 1. Types 1, 3 and 4 are reasonably consistent across nearly 6000\,s of data. Type 2, however, has a large variance in its metric values over time; this may be due to the ionospheric structures responsible, which are especially large compared to the MWA field of view, and may be drifting in and out of sight as the observations progress. Despite the large variance, the metric values indicate adverse observing conditions.

The data labelled `transition' in Fig.~\ref{fig:metric_variation_time} were collected on 2015-10-08. These data show metric values ranging from what would be classified as Type 4 down to Type 1. The ionosphere appears to be extremely active at the start of the observations, but steadily decreases in activity until approximately 3000\,s later, when the corresponding metric values are similar to our Type 1 data. This suggests that the ionosphere can slowly change from active to inactive on time scales of order an hour. While we do not witness any abrupt changes in ionospheric activity (order seconds or minutes), it seems that monitoring should be conducted at least hourly to ensure ideal observing conditions.

\subsection{Density of pierce points required to assess ionospheric activity}
\begin{figure}
  \includegraphics[width=\columnwidth]{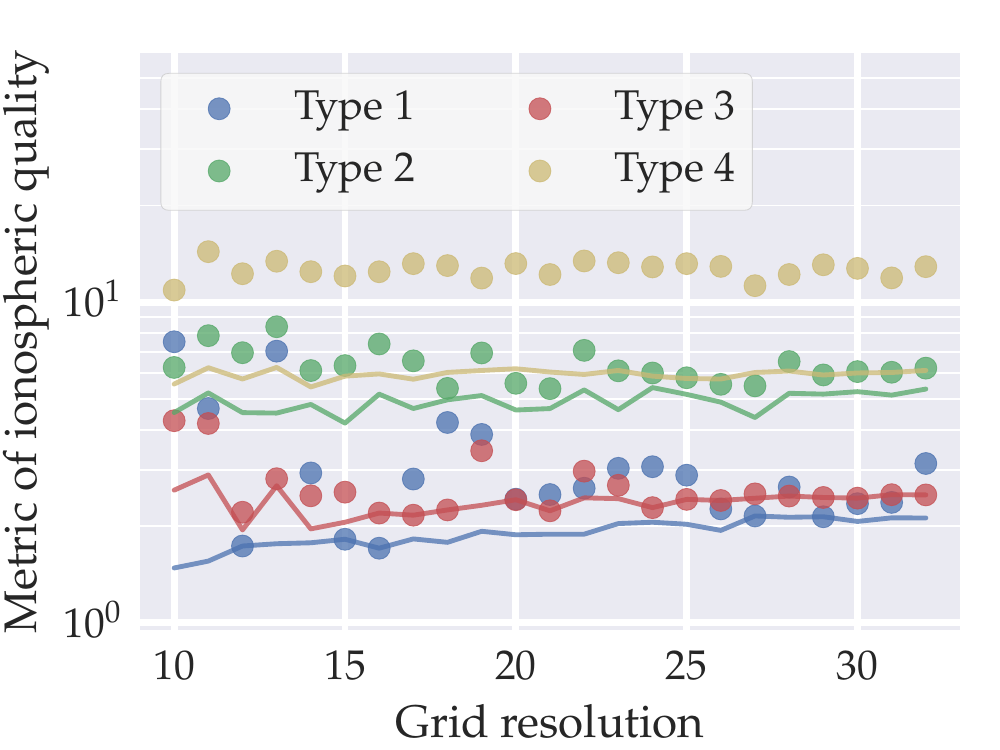}
  \caption{Scatter and line plot comparing the ionospheric quality metric for each of the four ionospheric types discussed in this work against pierce point density. The primary beam of the MWA (approximately 25-by-25 square degrees) was divided into evenly-sized bins, and the brightest calibrator (if any were available) in the bin was used in the final pierce point count. Using only these bright calibrators, the ionospheric quality metric was calculated and plotted with dots here. Type 1 and 3 data have large variances at small resolutions; when adjusting the calculation of the metric, the lines are obtained. The lines indicates that we can determine ionospheric activity even with few ($< 200$) pierce points.}
  \label{fig:metric_variation_pierce}
\end{figure}

In this work, each observation has been calibrated with 1000 sources, which provides a high density of pierce points within the primary beam of the MWA to diagnose ionospheric activity. However, using 1000 sources is computationally expensive; are we able to diagnose ionospheric activity with a smaller density of pierce points? In this section, we break the field of view (approximately 25-by-25 square degrees) into equally sized bins, and for each bin, choose the brightest source as our pierce point, gradually increasing the bin size. Results are shown in Fig.~\ref{fig:metric_variation_pierce}.

We find that the metric of ionospheric quality is generally consistent down to a grid resolution of 20 (20-by-20 boxes for a total of 400 pierce points). Fewer pierce points are less reliable for identifying Type 1 and 3 data, but seem robust for types 2 and 4. Type 1 and 3 data becomes erratic with fewer pierce points due to enhanced contribution of the PCA eigenvalue in the ionospheric quality metric; in order to work around this, the metric could be tweaked for this analysis to provide more consistent results. Indeed, by ignoring the magnitude of ionospheric offsets when performing a PCA, the metric value becomes smooth across all grid resolutions; see the lines in Fig.~\ref{fig:metric_variation_pierce}. This tweaked metric calculation allows for accurate determination of ionospheric activity, even with very few pierce points (100 at the low end). This is good news for low-frequency instruments that seek to alter their observing schedules in a short timeframe; an inexpensive calibration of data using few (order 300) pierce points can determine whether a sensitive project, such as the EoR, should be rescheduled early in an observing run.

With an understanding of how differing ionospheric types affect the desired measurable quantities (e.g. EoR power spectrum), we can refine this test to identify a fast, cheap way of identifying adverse ionospheric conditions. In addition, if only a few pierce points are used, it is possible to continuously characterise ionospheric activity in real-time.

\subsection{Comparison of ionospheric quality metric with planetary $K$ indices}
\label{ref:kp}
Geophysicists convert geomagnetic fluctuations measured in nanoTesla into $K$ indices, which take values between 0 and 9. $K$ indices provide a measure of activity in the geomagnetic field, typically caused by solar radiation \citep{menvielle91}. Thirteen geomagnetic observatories around the world record $K$ indices every 3~hours, which are then averaged to yield planetary $K$ indices ($K_p$). Due to dependence on geographic location, each observatory is subject to a different scale of geomagnetic fluctuation. Thus, the reported $K$ indices are adjusted from each observatory such that the historical average counts of reported $K$ indices matches all other observatories. Given that a geographic variation exists, and eleven of the thirteen geomagnetic observatories contributing to $K_p$ indices are in the northern hemisphere, far from the MRO, we compared data from two Australian observatories against the others, including one geographically close to the MRO (Gingin, the other being Canberra), but found no significant differences. In this work, we opt to use $K_p$ indices rather than data only from Australian geomagnetic observatories, because the averaged indices have higher resolution to their data. The data used in this work were taken from the National Centers for Environmental Information website\footnote{\url{https://www.ngdc.noaa.gov/stp/GEOMAG/kp_ap.html}}.

\begin{figure*}
  \includegraphics[width=\textwidth]{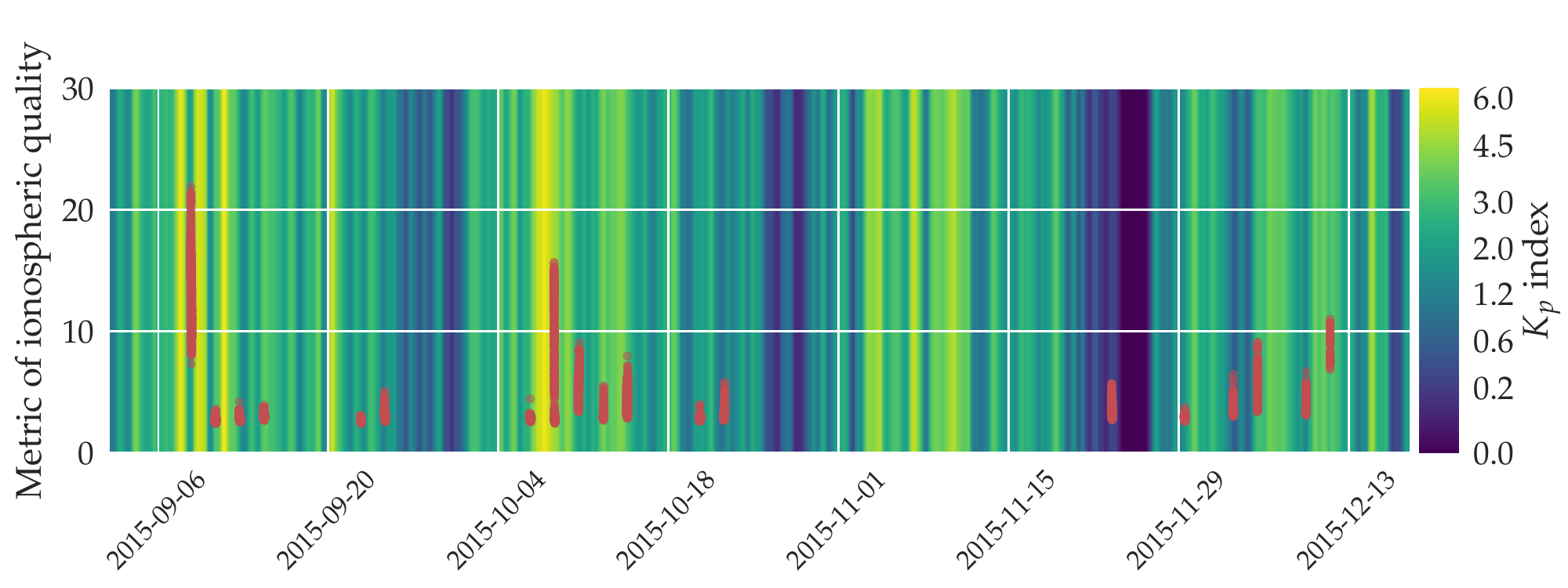}
  \caption{Ionospheric quality metric values for all observations detailed in this work plotted against time. The background colour represents the $K_p$ indices, with a single index plotted every 3~hours, but smoothed with a Hanning window of length 7. A spectrum of geomagnetic activity occurs over the dates of our ionospheric data, including four peaks in the $K_p$ indices occurring close in time to our observations. There is insufficient data to conclude a relationship.}
  \label{fig:metric_v_kp}
\end{figure*}

Fig.~\ref{fig:metric_v_kp} shows the ionospheric quality metric values for all observations detailed in this work, along with $K_p$ indices for every 3~hour interval. A number of peaks in the $K_p$ indices are visible, four of which occur close to our observations. The peak on the 7th of September precedes our most active ionospheric data by about 20~hours. We also have active ionospheric data after the October 7 peak, again lagging by approximately 20~hours. However, our data features inactive ionospheric data for the other two peaks in the $K_p$ indices on the 9th and 11th of September. The lags between $K_p$ index peaks and our data is in both cases approximately 27~hours. Perhaps the additional 7~hour difference is enough for conditions to calm after increased geomagnetic activity?

We conducted an analysis where the average $K_p$ index value some time before each metric value was compared against each other. When using the average $K_p$ index between 12 and 24 hours before our observations, we see our two most ionosphericly active nights along with a quiet one with $K_p$ indices greater than 5. Other time lags (24-36~hours, 12-36~hours) were also compared, and give similar results. However, because our results are so sparse, we are unable to draw any significant conclusions. If either of our two most ionosphericly active nights were not observed, then we would be inclined to believe there was no relation between our metric values and $K_p$ indices. Perhaps exceeding a certain threshold in the $K_p$ indices leads to an increased likelihood of enhanced ionospheric activity, but we would need more data to confirm this claim.

\section{Conclusions}
We have developed robust techniques for ionospheric characterisation with MWA EoR data, and have identified four prominent types of ionospheric activity. Fortunately for low-frequency astronomy, the ionosphere is mostly inactive. However, in the data used for this work, we occasionally observe structures in the STEC. In addition, it appears that the degree of ionospheric activity persists across the course of an evening, which allows low-frequency observatories to alter their observing schedules early in an evening.

From these data, we provide a summary of our conclusions:

(i) Ionospheric activity can be broadly categorised into four different types. The classification for each type is somewhat arbitrary, but serves to illustrate various properties of these different observations. Our category for observations without active ionospheric activity occupies approximately 74 per cent of all observations;

(ii) A metric of ionospheric quality, using only two statistics, may be simple enough to determine if the ionosphere is sufficiently quiet to undertake particular experiments, with both the MWA and future SKA on the Murchison site;

(iii) Two methods by which we can determine the diffractive scale of the ionosphere. The first, as detailed by \citet{mevius16}, analyses the ionospheric refraction applied to a single point source as it drifts across the sky over an observation. The second extends the concept from a single source to all available sources. The first method indicates larger diffractive scales than the second for turbulent conditions, possibly due to a biasing effect by introducing more sources and spatially-variant ionospheric properties, but both methods are broadly consistent. Inactive ionospheric data have diffractive scales of $6.3-7.3$\,km and $> 10$\,km for the first and second methods, respectively, while extremely active ionospheric data have diffractive scales of $2.9-4.5$\,km and $> 3.1$\,km for the first and second methods, respectively;

(vi) Temporal correlations of pierce points can determine the presence of ionospheric activity, and if any is present, show the scale and nature of the ionospheric structures;

(v) By using observations with 1000 pierce points, we have determined that as few as 100--200 pierce points are required to identify ionospheric activity;

(vi) The severity and type of ionospheric activity tends to remain constant across an entire evening of observations. However, in one of our nineteen nights of observations, we do see the ionospheric activity change from extremely active to inactive in one hour. This suggests that ionospheric activity should be monitored at least hourly;

(vii) Geomagnetic activity highlighted by $K_p$ indices do not appear to be correlated with our metric of ionospheric quality, but more data is required to confirm a relationship either way.

Finally, the degree, type, and frequency of ionospheric activity at the Murchison Radioastronomy Observatory studied in this work suggest that the core of SKA-Low is highly-calibratable, having a physical footprint of equivalent size to the MWA. Scintillation effects present in seemingly common Type 1 conditions can be calibrated \citep{vedantham16}, but in addition, due to the stability of ionospheric structures, it is possible that the effect of intense ionospheric activity can be largely subtracted from observed radio data.

\section*{Acknowledgements}
C. H. Jordan thanks Paul Hancock for providing data containing ionospheric activity early in the pipeline development, and discussing techniques on identifying ionospheric structure. Computation was aided with the \textsc{Python} libraries \textsc{numpy} \citep{numpy}, \textsc{scipy} \citep{scipy} and \textsc{astropy} \citep{astropy}. In addition, \textsc{matplotlib} was used to generate the figures presented in this publication \citep{matplotlib}. The Centre for All-Sky Astrophysics (CAASTRO) is an Australian Research Council Centre of Excellence, funded by grant CE110001020.  This research has made use of NASA's Astrophysics Data System. CMT is supported under the Australian Research Council's Discovery Early Career Researcher funding scheme (project number DE140100316). This work was supported by resources provided by the Pawsey Supercomputing Centre with funding from the Australian Government and the Government of Western Australia. We acknowledge the iVEC Petabyte Data Store, the Initiative in Innovative Computing and the CUDA Center for Excellence sponsored by NVIDIA at Harvard University, and the International Centre for Radio Astronomy Research (ICRAR), a Joint Venture of Curtin University and The University of Western Australia, funded by the Western Australian State government.



\bibliographystyle{mnras}
\bibliography{references}



\newpage
\appendix
\section{Surface reconstruction}
\label{app:reconstruction}
Interferometers are unable to measure the total path delay from the ionosphere, instead only measuring changes in the path delay, which manifest as refractive-like source offsets. These offsets directly probe the \emph{differential} STEC, rather than the scalar STEC field itself. However, the total content - quantified by the scalar STEC field - is a physically interesting quantity, and may be important for diagnosing the quality of a low-frequency observation. In this section, we present a novel method of deriving the scalar field from a catalogue of source offsets.

Let the scalar STEC field, i.e. the column density of electrons in the ionosphere along a given line-of-sight, $\vec{\theta}$, be denoted $\phi(\vec{\theta})$. We note that in general $\phi$ is also a function of time, but our calculations here are restricted to a specified time interval, and we omit it for clarity. In general then, a given source, $i$, will experience a 2D offset according to:
\begin{equation}
	\Delta \vec{\theta}_i \propto \nabla \phi(\vec{\theta}).
\end{equation}

Note that this assumes all antennas see through the same part of the ionosphere. Setting the constant of proportionality for now to $C$, we recognise that the individual offsets, $\vec{D}_i \equiv \Delta \vec{\theta}_i$, are drawn from a vector field:
\begin{equation}
\vec{D}(\vec{\theta}) = C \nabla \phi(\vec{\theta}).
\end{equation}

Supposing that our sample of offsets are dense enough to specify the vector field $\vec{D}$ to an adequate resolution - via some form of interpolation - our task is then to integrate $\vec{D}/C$ to obtain the scalar field $\phi$ (modulo an additive constant).

This task - the integration of a 2D vector field - is non-trivial. A first attempt may be made by prescribing a parametric form for $\phi$, performing the derivative analytically, and minimising the residual between the gradient field and the measured offsets. Clearly, the drawback of this method is its dependence on a good choice of parameterisation. Since \citet{loi15a} found sinusoidal structures in the ionosphere, it is tempting to use a generalised sinusoidal parameterisation. However, such structures are quite rare, and it is unknown how important they are for a typical observing night. Thus, we seek a non-parametric solution.

Following the same general arguments - specification of a field $\phi$ and subsequent minimisation of its residuals - a non-parametric method can be imagined. This method consists of specifying $N$ `nodes', $\vec{\theta}_j$ across the observed field, and assigning each a value $\phi_j$. The discrete field specified by these nodes is subsequently 2D-interpolated to yield an approximate continuous field for which the gradient can be uniquely calculated. 
One then proceeds again to minimise the residuals between this gradient and the measured offsets, where the parameters of the minimization are the $N$ values $\phi_j$. Tests of this method yielded good results in many cases, but ultimately we found that it was unreliably dependent on the resolution of the nodes chosen, with choices that were mismatched to the typical scales of structures in $\phi$ - both over- and under-dense - tending to result in numerical errors. In addition, this method is prohibitively expensive due to the minimisation process.

Despite this, the idea behind this method is optimally accurate. Two modifications to the process render it quite feasible: firstly, instead of determining derivatives on the grid via splines, the derivatives can be specified numerically using the familiar `central differences' formula, in matrix form. Secondly, instead of using a downhill-gradient method of minimization, the minimization can be done analytically, up to a necessarily numerical solution of a matrix equation.

This results in precisely the method described in \citet{harker15}, who show that the surface reconstruction problem is thus identical to the solution of a Sylvester Equation - something that can be achieved in $\mathcal{O}(N^3)$ time (where $N$ is the number of nodes in the surface grid). We outline this method for completeness in Appendix~\ref{app:harker}.

\section{The Harker-O'Leary method}
\label{app:harker}
This technique is detailed in \citet{harker15}, and we reproduce the basic outline here for completeness. They proceed to write down the numerical derivative of a function $f(x)$, discretised at given nodes, as a matrix equation:
\begin{equation}
	f'_j = D_{ij} f_i.
\end{equation}

The matrix $D$ is specified by the central difference formula of a given order (e.g. first-order central differences will have up to 3 contiguous non-zero elements in every row). 

Given a gridded surface, expressed as the matrix $Z$, the numerical derivatives in each direction are given by
\begin{align}
\nonumber	\frac{\partial Z}{\partial x} &= Z D^T \\
	\frac{\partial Z}{\partial y} &= D Z. 
\end{align}
Note that, for clarity, we have assumed that the grid-spacing of the surface is both uniform and the same in $x$ and $y$. The extension to a non-uniform and asymmetric grid is trivial. 

Following our intuitive minimisation method, the integration problem is to determine $Z$ such that for a measured gradient, $\hat{Z}$, we have:
\begin{equation}
	\hat{Z}_x \approx Z D^T\ \ , {\rm and\ \   } \hat{Z}_y \approx D Z.
\end{equation}

The value to minimise is thus:
\begin{equation}
	|| Z D^T - \hat{Z}_x||^2 + ||DZ - \hat{Z}_y||^2,
\end{equation}
yielding the matrix equation:
\begin{equation}
	D^T D Z + ZD_xD - D_y^T\hat{Z}_y - \hat{Z}_xD = 0.
\end{equation}

This final equation is a well-known form, termed a Sylvester Equation, and must be numerically solved. However, solutions have been well-studied, and the most efficient are $\mathcal{O}(N^3)$, where the number of nodes in $Z$ is $N$. This greatly improves on previous state-of-the-art reconstruction methods which only offer $\mathcal{O}(N^{6})$. Indeed the computation time required for reconstruction in this paper is short (less than one second on an Intel i7-6700K processor). 

We note that the preceding derivation is valid when the noise at each node is independently drawn from a single normal distribution. In general, this may not be a good approximation; some offsets will be more certain than others. \citet{harker15} provide generalisations of the method to accommodate several forms of regularisation, including weighted-least-squares. We do not use these generalisations in this paper, leaving their exploration for future work.

\subsection{Implementation of the Harker-O'Leary method}
The Harker-O'Leary method requires as input an observed gridded vector field $\vec{D}$. However, we begin with a set of $n$ observed offsets at arbitrary positions. To obtain an estimation of the gridded field, we perform 2D linear interpolation onto a set of $N$ grid-nodes. Our choice of linear interpolation, as opposed to some higher-order variant, is based on a desired preservation of noise characteristics. Assuming that the noise in each offset is independent and Gaussian, then the variance at some intervening point is the distance-weighted sum of the variances, which is provided by linear interpolation.

The choice of $N$ is also important. Clearly as few as possible are desired, but enough to resolve the structures at hand. We choose to use $N \sim n$ in the reconstructions in this paper.

To solve the Sylvester Equation, we have translated the MATLAB code provided by \citet{harker15} into \textsc{Python}, and included this alongside \textsc{cthulhu}\footnote{\url{https://github.com/cjordan/pyGrad2Surf}}.

This method has proven remarkably effective at reconstructing sparsely-sampled simulated surfaces; see Appendix~\ref{app:g2s} for a demonstration, as well as for a comparison of reconstructions for differing numbers of pierce points.

\section{Simulations of the Harker-O'Leary method with variable pierce points}
\label{app:g2s}
Here, we test the method used to reconstruct STEC scalar fields by generating a scalar field (our STEC), then simulating its reconstruction with a variable number of pierce points. The input STEC scalar field is derived from a Gaussian-distributed noise map with an isotropic power-law power spectrum with index $-$3. This results in scalar fields with `clumpy' large-scale structure, but with random small-scale irregularities. Our pierce points are randomly but uniformly placed over the STEC, and the gradient of the TEC is determined and recorded for each pierce point. Using the positions of the pierce points and gradients at the pierce points, we use the same reconstruction method described in the text.

\begin{figure*}
  \includegraphics[height=0.9\textheight]{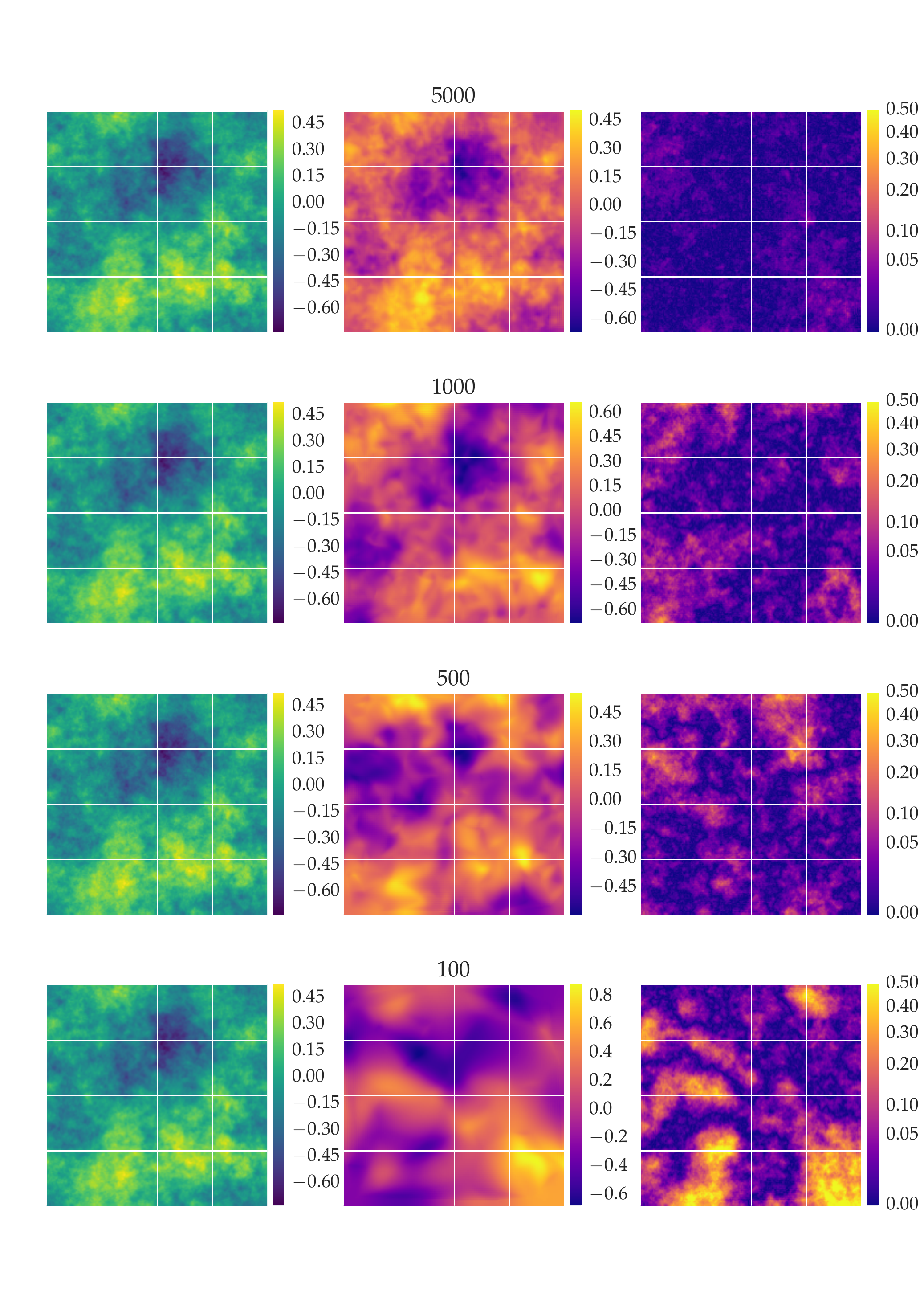}
  \caption{Reconstructions of a generated TEC scalar field using the Harker-O'Leary method. Each row of plots contains the generated TEC, the reconstructed TEC, and the residual map. The pierce points are uniformly distributed across the generated TEC, and the number of pierce points used for reconstruction is indicated above each reconstructed TEC. The residual map is generated by taking the Euclidean norm between corresponding pixels.}
  \label{fig:appendix_g2s_showcase}
\end{figure*}

Fig.~\ref{fig:appendix_g2s_showcase} shows the same input TEC scalar field reconstructed with 5000, 1000, 500 and 100 pierce points. The third panel of each row is the residual, calculated by Euclidean norm between corresponding pixels. The reconstructed TEC generated with 5000 pierce points has visually little difference from the input TEC, confirmed by small residual values. The reconstructed TEC derived from 1000 pierce points has less small-scale features, but appears to map large-scale features accurately, and has dissimilar structure in the residual map. Reconstructed TECs with 500 and 100 pierce points have clear departures from their input TECs, and have differences in the overall scale of pixel values. Note that input TEC scalar fields used a size of 200-by-200 pixels.

\begin{figure*}
  \includegraphics[width=\textwidth]{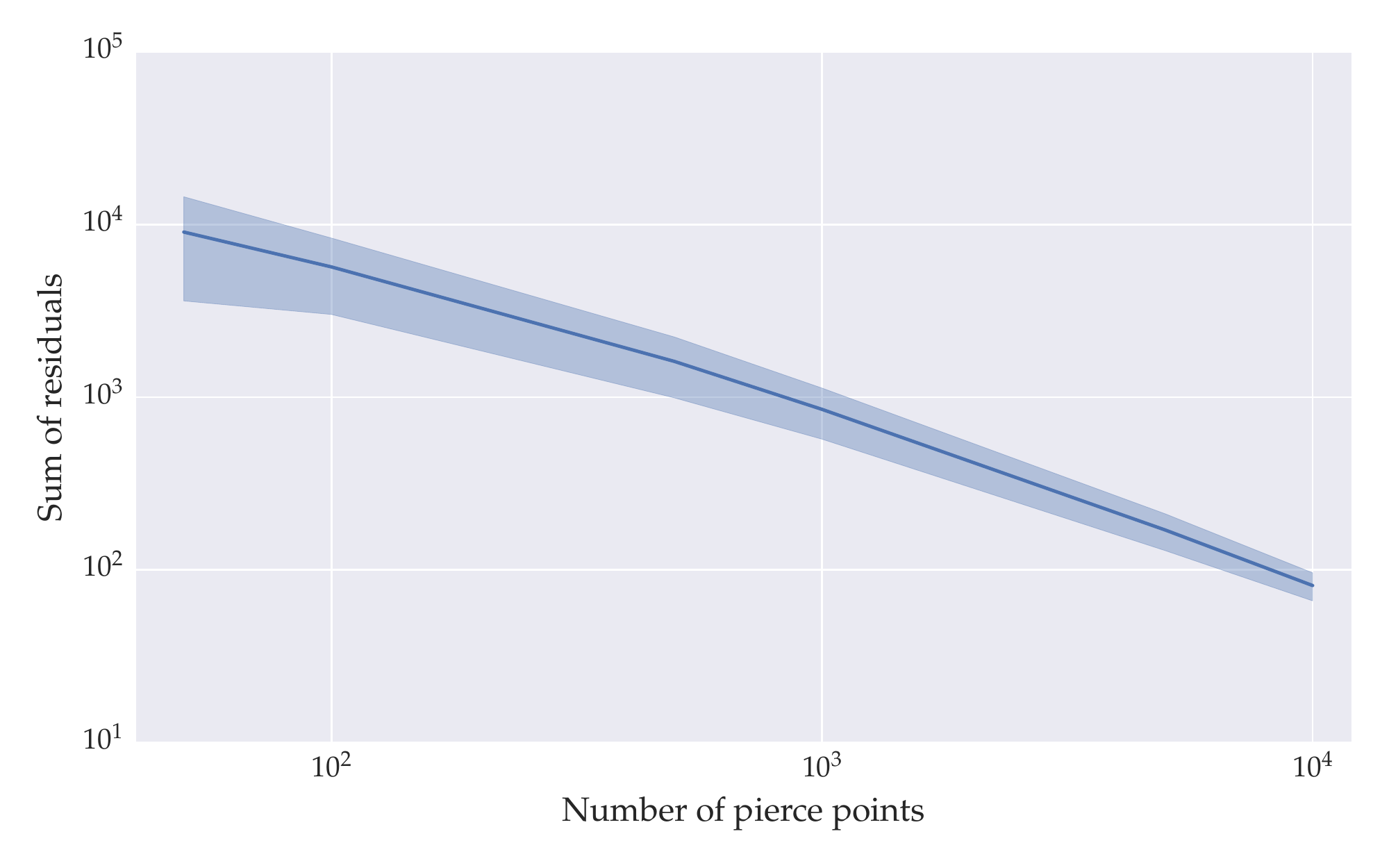}
  \caption{Results of a Monte Carlo simulation of TEC scalar field reconstructions. One thousand simulated TEC scalar fields were each analysed with 50, 100, 500, 1000, 5000 and 10000 pierce points. The Euclidean norm is calculated for each input TEC and reconstructed TEC, and the sum of all norms is reported in the figure, as well as its 1$\sigma$ boundaries. Note that the size of the scalar fields was 200-by-200; the number of pixels will directly affect the sum of residuals calculated. The results approximate a power law.}
  \label{fig:appendix_g2s_error}
\end{figure*}

Fig.~\ref{fig:appendix_g2s_error} shows the results of a Monte Carlo simulation, generalising the results shown in Fig.~\ref{fig:appendix_g2s_showcase}. A range of pierce-point counts are selected over 1000 input TECs, and the sum over all pixels of the resulting residual map was recorded. Using the sums, we have plotted the mean and standard deviation against pierce-point counts. The results approximate a straight line in log space; an order of magnitude more pierce points will approximately reduce the resulting error by an order of magnitude.

\bsp	
\label{lastpage}
\end{document}